\documentstyle[aps,epsfig,preprint]{revtex}

\begin{document}
\tighten
\draft
\preprint{
\vbox{
\hbox{ADP-98-20/T296}
\hbox{FZ-IKP(TH)-1998-13}
\hbox{September 1998}
}}

\title{Dynamics of Light Antiquarks in the Proton}

\author{W. Melnitchouk}
\address{Institut fuer Kernphysik,
	Forschungszentrum Juelich,
	D-52425 Juelich, Germany	\\
	{\tt w.melnitchouk@fz-juelich.de}}
\author{J. Speth}
\address{Institut fuer Kernphysik,
	Forschungszentrum Juelich,
	D-52425 Juelich, Germany	\\
	{\tt j.speth@fz-juelich.de}}
\author{A.W. Thomas}
\address{Department of Physics and Mathematical Physics,
	and Special Research Centre for the Subatomic
	Structure of Matter,
	University of Adelaide,
	Adelaide 5005, Australia	\\
	{\tt athomas@physics.adelaide.edu.au}}

\maketitle

\begin{abstract}
We present a comprehensive analysis of the recent data from
the E866 experiment at Fermilab on Drell-Yan production in
$pD$ and $pp$ collisions, which indicates a non-trivial
$x$-dependence for the asymmetry between $\bar u$ and $\bar d$
quark distributions in the proton.
The relatively fast decrease of the asymmetry at large $x$
suggests the important role played by the chiral structure
of the nucleon, in particular the $\pi N$ and $\pi\Delta$
components of the nucleon wave function.
At small $x$ the data require an additional non-chiral component,
which may be attributed to the Pauli exclusion principle, as first
suggested by Field and Feynman.
\end{abstract}

\section{Introduction}

The recent Drell-Yan experiment by the E866/NuSea Collaboration at
Fermilab \cite{E866} provides the best information yet on the detailed
structure of the light antiquark sea of the proton.
Although previous experiments by the New Muon Collaboration \cite{NMC}
on the difference between the proton and neutron structure functions
established that an asymmetry in the sea exists, they yielded direct
information only on the first moment of the antiquark asymmetry.
The earlier Drell-Yan experiment by the NA51 Collaboration
at CERN \cite{NA51} measured the up and down antiquark ratio,
though at zero rapidity \cite{ES}, but in order to improve the
statistical accuracy the data were binned to a single value of
Bjorken-$x$.
The E866 experiment, on the other hand, has for the first time
mapped out the shape of the $\bar d / \bar u$ ratio over a large
range of $x$, $0.02 < x < 0.345$.

The relatively large asymmetry found in these experiments implies the
presence of non-trivial dynamics in the proton sea which does not have
a perturbative QCD origin.
{}From the symmetry properties of QCD, we know that one source of
non-perturbative quark-antiquark pairs is the pion cloud associated
with dynamical chiral symmetry breaking.
The novel and unexpected feature of the E866 data is that the
$\bar d/\bar u$ asymmetry peaks at rather small values of $x$,
and drops quite rapidly with increasing $x$, approximately like
$(1-x)^n$ with $n \sim 10$.
This behavior had not been anticipated in global parameterizations
of data, and is softer than pion cloud models of the nucleon would
generally predict.
While the former is just an artifact of an extrapolation of a
parametric function into an unmeasured region, with no physics
implications, the reason for the latter is the rather hard valence
antiquark distribution in the pion, $\bar q_v^{\pi} \sim (1-x)^n$
with $n \sim 1$.
Although this $x$ dependence is softened somewhat by the probability
of finding the pion in the nucleon in the first place, the resulting
convoluted distribution still does not die off as rapidly, at large
$x$, as the new E866 data would suggest.

Our analysis suggests that a quantitative description of the entire
region of $x$ covered in the experiment requires a delicate balance
between several competing mechanisms, which leads us to speculate
about the possibility of a two-phase picture of the non-perturbative
sea of the nucleon.
At larger $x$, the dynamics of the pion cloud of the nucleon come
to the fore, with deep-inelastic scattering from the $\pi N$
component of the nucleon wave function providing the bulk of the
$\bar d-\bar u$ asymmetry.
On the other hand, the important role of the $\Delta$ isobar in
nuclear physics has been known for a long time, and it proves to
be of some importance in this case too.
If a part of the $\pi\Delta$ distribution happens to be harder
than the $\pi N$, there would be cancellation of some of the
$\bar d$ excess at large $x$.
Such a piece does indeed arise in the light-cone formulation of
the meson-cloud model, provided the $\pi N \Delta$ form factor,
parameterized for example by a dipole form, is harder than the
$\pi NN$ form factor \cite{MTV} --- something which is consistent
with the measured difference between the nucleon and $N\Delta$
transition axial form factors.

To be consistent with the measured sum, $\bar u + \bar d$, it is
known that both the $\pi N N$ and $\pi N \Delta$ form factors need
to be relatively soft \cite{AWT83}.
The best fit, within the pion cloud framework, to the data on both
the sum and difference of $\bar u$ and $\bar d$ at large $x$ accounts
for around half of the integrated asymmetry, leaving room for possible
other, non-pionic, mechanisms to provide the missing strength at
smaller $x$.
(While the usual discussions of the pionic contribution focus on the
valence quarks in the pion, there is also some theoretical argument
for a $\bar{d} - \bar{u}$ asymmetry in the sea of the pion.
We also estimate how this might affect the analysis.)

It should be noted that, aside from the flavor-asymmetric $\pi N$
and $\pi \Delta$ components of the nucleon, there is no {\em a priori}
reason why the `bare' (non--pion-dressed) nucleon state itself cannot
have an intrinsic asymmetric sea associated with it.
In fact, this is actually what is expected from the Pauli exclusion
principle, as anticipated long ago by Field and Feynman \cite{FF}
on the simple basis that the $u$ and $d$ valence quark sectors are
unequally populated in the proton ground state.
Although more difficult to  estimate model-independently, the contribution
to the $\bar d-\bar u$ difference from antisymmetrization has been
calculated within a non-perturbative model of the nucleon \cite{BAG}.
Along the lines of the model estimates, we find that the effects of
antisymmetrization are most relevant at small $x$, with normalization
such that they can account for most of the remaining half of the
integrated asymmetry.
Indeed, our analysis suggests that the best fit to the E866 data
is obtained when both mechanisms play a role, consistent with the
conclusions of earlier analyses \cite{SST,MTS} of the NMC data on
the proton--neutron structure function difference.

In the following, we firstly outline in Section II the experimental
status of the $\bar d / \bar u$ asymmetry, including a comparison
of the Drell-Yan data with deep-inelastic scattering data.
In Section III the question of the possible origin of the asymmetry
is addressed, in the form of the chiral structure of the nucleon
and the associated pion cloud, while Section IV deals with the role
of the Pauli exclusion principle as a source of flavor asymmetry.
In Section V we summarize our conclusions.

\section{The Light Antiquark Asymmetry}

The E866/NuSea Collaboration measured $\mu^+\mu^-$ Drell-Yan pairs
produced in $pp$ and $pD$ collisions.
In the parton model the Drell-Yan cross section is proportional to:
\begin{eqnarray}
\sigma^{ph}
&\propto& \sum_q e_q^2
\left( q^p(x_1)\ \bar q^h(x_2) + \bar q^p(x_1)\ q^h(x_2) \right),
\end{eqnarray}
where $h = p$ or $D$, and $x_1$ and $x_2$ are the light-cone momentum
fractions carried by partons in the projectile and target hadron,
respectively.

Assuming that the deuteron is composed of two bound nucleons,
and utilizing isospin symmetry ($u^p = d^n$, etc.), in the limit
$x_1 \gg x_2$ (in which $\bar q(x_1) \ll q(x_1)$) the ratio of the
deuteron to proton cross sections can be written:
\begin{eqnarray}
\label{Rful}
\left. { \sigma^{pD} \over 2 \sigma^{pp} } \right|_{x_1 \gg x_2}
&=& {1 \over 2}
\left( {\widetilde{\bar u}(x_2) \over \bar u(x_2)}
     + {\widetilde{\bar d}(x_2) \over \bar u(x_2)}
\right)
{ 4 + d(x_1)/u(x_1) \over
  4 + d(x_1)/u(x_1) \cdot \bar d(x_2)/\bar u(x_2) }\ \ ,
\end{eqnarray}
where $\widetilde{\bar q}$ is the antiquark distribution in the
bound proton.
Neglecting relativistic and nucleon off-shell effects, this can
be approximated by a convolution of the antiquark distribution
in the proton with the proton distribution function in the deuteron
\cite{SMEAR,SMEAR2},
\begin{eqnarray}
\widetilde{\bar q}(x) &\approx&
\int_x {dz \over z} f_{N/D}(z)\ \bar q(x/z),
\end{eqnarray}
where $f_{N/D}(z)$ is the distribution of nucleons in the deuteron
with light-cone momentum fraction $z$.
In practice we use the function $f_{N/D}$ from Ref.\cite{SMEAR},
where it is given in terms of a realistic deuteron wave function
that has been constrained to reproduce the static deuteron
properties and nucleon--nucleon phase shifts.

In the absence of nuclear effects, $\widetilde{\bar q} = \bar q$,
one would have:
\begin{eqnarray}
\label{nonuc}
{ \sigma^{pD} \over 2 \sigma^{pp} }
&=& {1 \over 2}
\left( 1 + { \bar d(x_2) \over \bar u(x_2) } \right)
{ 4 + d(x_1)/u(x_1) \over
  4 + d(x_1)/u(x_1) \cdot \bar d(x_2)/\bar u(x_2) }\ ,
\end{eqnarray}
so that the ratio would be unity if $\bar d = \bar u$.
On the other hand, we know that nuclear shadowing exists in the
deuteron at small $x$ (see \cite{MTD} and references within),
and at large $x$ nuclear binding and Fermi motion effects come
into prominence \cite{NP}.
Since the bulk of the effect is observed outside the very small
$x$ region, shadowing will not be important in these data.
However, Fermi smearing is potentially more relevant, as what enters
in the parentheses in Eq.(\ref{Rful}) is the ratio of smeared to
unsmeared {\em sea} quark distributions, for which smearing effects
should come into play at much smaller $x$ than for the total structure
function \cite{SMEAR}.
In Fig.~1 we show the ratio of the antiquark distributions
in a proton bound in the deuteron to that in a free proton,
$\widetilde{\bar q} / \bar q$.
Parameterizing the antiquarks for illustration purposes at large $x$
by a simple $\bar q \sim (1-x)^n$ form, with $n=5, 7$ and 10,
the ratio is seen to rise rapidly above $x \sim 0.4$, though in
the measured region it only deviates from unity by a few percent.
{}From this one can conclude that Eq.(\ref{nonuc}) should be a
reasonable approximation to Eq.(\ref{Rful}).

In the extreme large-$x_1$ limit, where $d(x_1) \ll u(x_1)$,
the cross section ratio would directly give $\bar d/\bar u$:
\begin{eqnarray}
\label{Rapprox}
\left. { \sigma^{pD} \over 2 \sigma^{pp} } \right|_{d(x_1) \ll u(x_1)}
&\longrightarrow& {1 \over 2} \left( 1 + {\bar d(x_2) \over \bar u(x_2)}
\right).
\end{eqnarray}
For the E866 data the criterion $x_1 \gg x_2$ is not always satisfied,
however, so that Eq.(\ref{Rapprox}) gives only an indication of the
sensitivity of the Drell-Yan cross section to $\bar d/\bar u$, and in
practice the full parton model cross section is used together with an
iterative procedure in which the valence and total sea distributions
are assumed known and the extracted $\bar d/\bar u$ ratio adjusted to
fit the data \cite{E866}.
The resulting $\bar d/\bar u$ ratio is shown in Fig.~2, where the
asymmetry is found to peak at relatively small $x$, $x \sim 0.15$,
dropping rapidly to unity by $x \sim 0.3$.
Also shown for comparison is the NA51 data point \cite{NA51},
extracted from the Drell-Yan $pp$ and $pn$ asymmetry at $x=0.18$,
which lies slightly above the E866 data.
The two fits are from the CTEQ4 \cite{CTEQ4} parameterization
(dashed line), which has the asymmetry extending out to larger $x$,
and the more recent MRS98 analysis \cite{MRST} (dotted line),
which included the E866 data.

{}From the $\bar d/\bar u$ ratio the E866/NuSea Collaboration further
extract the difference, $\Delta \equiv \bar d - \bar u$, assuming the
sum $\bar d + \bar u$ from the CTEQ4 fit,
\begin{eqnarray}
\label{DeltaRSum}
\Delta
&=& \left( { \bar d/ \bar u - 1 \over \bar d/\bar u + 1 } \right)
    \left( \bar d + \bar u \right)_{\rm CTEQ4}.
\end{eqnarray}
The resulting points are shown in Fig.~3, again in comparison with the
CTEQ4 and MRS98 parameterizations.
It should be noted, however, that the sea quark distribution in these
fits is not very well determined at large $x$.
In particular, a larger total sea at $x \sim$ 0.2--0.3 would result
in a larger asymmetry $\Delta$ in the region just where the E866 data
appear to drop rapidly to zero.
Further data from the E866 experiment on the total antiquark
distribution at large $x$ should help to clarify the issue.
The Drell-Yan data can also be compared with the
proton--neutron\footnote{Note that the NMC extracted the
neutron structure function from proton and deuteron data assuming
$F_2^n = F_2^D - F_2^p$.}
structure function difference measured previously by the New Muon
Collaboration \cite{NMC}, if one assumes that the valence quark
distributions in the proton are known.
In this case the antiquark asymmetry can be written:
\begin{eqnarray}
\bar d(x) - \bar u(x)
&=& {1 \over 2}  \left( u_V(x) - d_V(x) \right)
 -  {3 \over 2x} \left( F_2^p(x) - F_2^n(x) \right)_{\rm NMC}.
\end{eqnarray}
The $\bar d - \bar u$ difference extracted from the NMC data
is shown in Fig.~3 using both the CTEQ4 (open circles) and MRS98
(diamonds) fits to the valence quark distributions.
The NMC values appear to lie consistently above the E866 data
for $x$ above $\sim 0.1$, although there is some sensitivity
to the choice of valence quark parameterization.
Not surprisingly, the E866 integrated value (after extrapolating
down to $x=0$ and up to $x=1$) is found to be:
$\Delta^{\rm E866} = \int_0^1 dx (\bar d - \bar u) = 0.100 \pm 0.018$
\cite{E866THEORY}, somewhat smaller than the NMC value, which is
$\Delta^{\rm NMC} = 0.148 \pm 0.039$ \cite{NMC}, although still
consistent within errors.
This difference will be further enhanced if one corrects the
NMC data for shadowing in the deuteron, omission of which
underestimates the violation of the Gottfried sum rule \cite{MTD}.

The observation of a large asymmetry between $\bar u$ and $\bar d$,
now both at CERN and Fermilab, provides theorists with a challenge
to better understand the internal, non-perturbative structure of
the nucleon, as the asymmetry due to perturbative effects is known
to be very small \cite{RS}.
In the next section we consider the possible origin of this asymmetry,
in the form of the non-perturbative chiral structure of the nucleon.

\section{Chiral Symmetry and the Meson Cloud}

The simplest and most obvious source of a non-perturbative asymmetry
in the light quark sea is the chiral structure of QCD.
{}From numerous studies in low energy physics, including chiral
perturbation theory, pions are known to play a crucial role in the
structure and dynamics of the nucleon.
However, there is no reason why the long-range tail of the nucleon
should not also play a role at higher energies.
This was first alluded to by Sullivan \cite{SULL}, who argued that
deep-inelastic scattering from the pion cloud of the nucleon is a
scaling contribution to the nucleon structure function.
Indeed, expectations for the ratio of nuclear to nucleon structure
functions, based on arguments that nuclear scales are much smaller
than typical deep-inelastic scales, and therefore irrelevant, were
proved to be dramatically wrong by the observation of the nuclear
EMC effect \cite{EMCEFFECT}.

As pointed out by Thomas \cite{AWT83}, if the proton's wave function
contains an explicit $\pi^+ n$ Fock state component, a deep-inelastic
probe scattering from the virtual $\pi^+$, which contains a valence
$\bar d$ quark, will automatically lead to a $\bar d$ excess
{\em in the proton}.
This is the essential physical idea behind these expectations,
and has been used to address not only the $\bar d / \bar u$
asymmetry \cite{MTV,SST,MTS,HM,KL,HSB,HSS,EHQ,KFS,REVIEW},
but also SU(3) flavor symmetry breaking in the proton sea
\cite{AWT83}, as well as asymmetries in the strange \cite{MM}
and heavier flavor sectors \cite{CHARM,PNNDB}.
In recent years this picture has been refined and elaborated with
inclusion of additional meson and baryon states \cite{MTV,HSS},
and constraints put on many of the model parameters by comparisons
of the predictions of the model with other processes \cite{ZOL,NSZ}.

The basic hypothesis of the meson cloud model is that the physical
nucleon state can be expanded (in an infinite momentum frame (IMF)
and in the one-meson approximation) in a series involving bare
nucleon and two-particle, meson--baryon states.
The essential ingredients are the meson--baryon distribution functions,
$f_{MB}(y)$, which give the probability to find a meson, $M$, in the
nucleon carrying a fraction $y$ of the nucleon's light-cone momentum.
As discussed at length in the literature, for these functions to
have the correct probabilistic interpretation in the IMF, they
must be related to the distributions of baryons in the nucleon,
$f_{BM}(y)$, via:
\begin{eqnarray}
\label{fysym}
f_{MB}(y) &=& f_{BM}(1-y).
\end{eqnarray}
This constraint can be verified easily in the IMF, but is not entirely
clear in covariant formulations (see Refs.\cite{MTV,HSS,REVIEW} for
further discussion on this point).
The IMF treatment has the additional advantage that the meson and
baryon are on-mass-shell and so one has no ambiguities associated
with the possible off-mass-shell behavior of their structure functions
that are encountered in the covariant treatments.

The contribution to the antiquark distribution in the proton,
$\delta^{(MB)} \bar q$, can then be written in the IMF as a
convolution of the meson distribution function and the antiquark
distribution in the (on-mass-shell) pion:
\begin{eqnarray}
\label{conv}
\delta^{(M B)} \bar q(x)
&=& \int_x^1 {dy \over y} f_{MB}(y)\ \bar q^M (x/y).
\end{eqnarray}
Note that this is the leading contribution to the antiquark
distribution, and is independent of the model of the bare
nucleon states.

In earlier studies it was found, not surprisingly, that pions are
the most important mesons, and that the dominant contributions
are those associated with the $\pi N$ component of the proton's
wave function.
The distribution of pions with a recoiling nucleon is given by
\cite{MTV,REVIEW,ZOL}:
\begin{eqnarray}
\label{fypin}
f_{\pi N}(y)
&=&
{ 3 g^2_{\pi N N} \over 16 \pi^2 }\
\int_0^{\infty}\ { dk_T^2 \over (1-y) }
\frac{ {\cal F}_{\pi N}^2(s_{\pi N}) }
     { y\ (M^2 - s_{\pi N})^2 }
\left( { k_T^2 + y^2 M^2 \over 1-y } \right),
\end{eqnarray}
so that $f_{\pi^+ n} = 2 f_{\pi^0 p} = (2/3) f_{\pi N}$
for the respective charge states.
The invariant mass squared of the $\pi N$ system is given by
$s_{\pi N} = (k^2_T + m_{\pi}^2)/y + (k^2_T + M^2)/(1-y)$,
and for the functional form of the $\pi NN$ vertex form factor
${\cal F}_{\pi N}(s_{\pi N})$ we take a simple dipole
parameterization:
\begin{eqnarray}
\label{ff}
{\cal F}_{\pi N}(s_{\pi N})
&=& \left( { \Lambda_{\pi N}^2 + M^2
       \over \Lambda_{\pi N}^2 + s_{\pi N} }
    \right)^2,
\end{eqnarray}
normalized so that the coupling constant $g_{\pi N N}$ has its
standard value (= 13.07) at the pole (${\cal F}(M^2)~=~1$).
The symmetry relations (\ref{fysym}) are automatically guaranteed
with this type of form factor, whereas in the earlier covariant
formulations \cite{SST,MTS,HM,KL,HSB}, with $t$-dependent form
factors, this could not be achieved.
Note that the E866 group also utilized the formulation in terms of
$t$-dependent form factors in their recent theoretical analysis
\cite{E866THEORY} of the Drell-Yan data.

The antiquark distribution in the pion has been measured in $\pi N$
Drell-Yan experiments by the E615 Collaboration at Fermilab \cite{E615}
and by the NA10 \cite{NA10} and NA3 \cite{NA3} Collaborations at CERN.
These have been parameterized in next-to-leading order analyses in
Refs.\cite{GRSPI,SMRS}.
Unless stated otherwise, we use the valence part of the pion's
antiquark distribution from Ref.\cite{GRSPI} throughout this analysis.

Because the meson cloud model is a model of part of the
(non-perturbative) sea, it can only be reliably applied to
describing the non-singlet $\bar d - \bar u$ distribution.
In Fig.~4 we show the calculated difference arising from the $\pi N$
component of the proton's wave function for two different cut-off
masses, $\Lambda_{\pi N} = 1$~GeV (dashed) and 1.5~GeV (solid),
giving average multiplicities
$\langle n \rangle_{\pi N} \equiv \int_0^1 dy f_{\pi N}(y) = 13\%$
and 26\%, respectively.
With the latter one has excellent agreement at intermediate $x$,
$x \alt 0.2$, while somewhat overestimating the data at the
larger $x$ values.
The excess at large $x$ is less severe for the smaller cut-off.
However, the strength of the contribution in that case is too small
at lower $x$.

To reconstruct the ratio from the calculated difference,
we assume, following E866, that the total $\bar d + \bar u$
is given by the CTEQ4 parameterization \cite{CTEQ4}, and
invert Eq.(\ref{DeltaRSum}).
The resulting ratio is plotted in Fig.~5.
At small $x$ the agreement with the data for the larger cut-off
is clearly excellent, but at larger $x$ the calculation does
not follow the downward trend suggested by the data, similar
to the CTEQ4 parameterization in Fig.~2.

Both Figs.~4 and 5 suggest that the excess of $\bar d$ over
$\bar u$ is too strong for $x \agt 0.2$, and that a mechanism
which suppresses or cancels this excess could be responsible
for the behavior seen in the data.
One way to obtain such a suppression would be if the pion
structure function were softer.
Actually, perturbative QCD suggests that the leading twist part
of $F_2^{\pi}$ should behave like $(1-x)^2$ \cite{FJ} at large $x$
(see also Ref.\cite{BBS}), although the Drell-Yan data
\cite{E615,NA10,NA3} show that it is closer to $(1-x)$,
even with the inclusion of higher twist contributions.

Within the pion cloud model another way that some of the
cancellation can be understood is through the $\Delta$ isobar.
Although the $\pi \Delta$ component should be smaller in
magnitude than the $\pi N$, if some part of the $\pi\Delta$
distribution were to be harder than the $\pi N$ it would allow
for some cancellation of the large-$x$ excess, while preserving
more of the asymmetry at smaller $x$.
In fact, qualitatively such behavior is exactly what is
seen in the model.

In the IMF, the pion distribution function with a recoil $\Delta$
is given by \cite{MTV,REVIEW,ZOL}:
\begin{eqnarray}
\label{fypid}
f_{\pi \Delta}(y)
&=& {2 g_{\pi N \Delta}^2 \over 16 \pi^2 }
\int_0^{\infty}\ { dk_T^2 \over (1-y) }
{ {\cal F}^2_{\pi \Delta}(s_{\pi\Delta})
  \over y\ (s_{\pi\Delta} - M^2)^2 }		\nonumber\\
& & \hspace*{2cm} \times
{ \left[ k_T^2 + (M_{\Delta} - (1-y) M)^2 \right]
  \left[ k_T^2 + (M_{\Delta} + (1-y) M)^2 \right]^2
\over 6\ M_{\Delta}^2\ (1-y)^3},
\end{eqnarray}
where $s_{\pi\Delta}$ is the $\pi \Delta$ invariant mass squared
and we take the same functional form (c.f. Eq.(\ref{ff})) for the
$\pi N \Delta$ form factor as for $\pi N N$.
The different $\pi \Delta$ charge states are obtained from
Eq.(\ref{fypid}) via
$f_{\pi^+ \Delta^0} = (1/2) f_{\pi^0 \Delta^+}
                    = (1/3) f_{\pi^- \Delta^{++}}
                    = (1/6) f_{\pi \Delta}$.
The $\pi N \Delta$ coupling constant is defined by:
\begin{eqnarray}
\langle N \pi | H_{int} | \Delta \rangle
= g_{\pi N \Delta}\ C^{t_N\ t_\pi\ t_\Delta}_{1/2\ 1\ 3/2}\
	\bar u(p_N,s_N)\ (p_N^{\alpha} - p_{\Delta}^{\alpha})\
	u_{\alpha}(p_{\Delta},s_{\Delta}),
\end{eqnarray}
with $u_{\alpha}$ the Rarita-Schwinger spinor-vector.
The value of $g_{\pi N \Delta}$ can be related to the $\pi N N$
coupling constant via SU(6) symmetry,
$g_{\pi N \Delta} = (6\sqrt{2}/5) f_{\pi N N}/m_{\pi}
                  \approx 11.8$ GeV$^{-1}$
(with $f_{\pi N N} = (m_{\pi}/2 M)\ g_{\pi N N}$).
Note that in Ref.\cite{PUMPLIN} (see also \cite{KL}) the
$\pi N \Delta$ coupling was extracted from the width of the
decay $\Delta \rightarrow N \pi$, giving a somewhat larger
value $g_{\pi N \Delta} \simeq 15.9$ GeV$^{-1}$ compared
with the SU(6) value.
However, from numerous studies \cite{PI_N} of $\pi N$ scattering
in the $\Delta$ resonance region, it is known that
up to 50\% of the $\Delta$ width comes from $\pi N$ rescattering
(for example, through diagrams of the Chew-Low type),
so that it would be inappropriate to acribe the entire
width to the tree level process in determining $g_{\pi N \Delta}$.
We expect the SU(6) value for the $\pi N \Delta$ coupling
to be accurate to within $\sim$ 10--20\%.

In Fig.~6 we show the $\pi\Delta$ distribution function as a
function of $y$, compared with the $\pi N$ distribution (\ref{fypin}).
The latter is calculated with a form factor cut-off, $\Lambda_{\pi N}$,
of 1~GeV, while the former has $\Lambda_{\pi\Delta}=1.3$~GeV,
which gives $\langle n \rangle_{\pi\Delta} = 11\%$.
The $\pi\Delta$ distribution is somewhat broader than the $\pi N$,
peaking at slightly larger $y$ (0.3 c.f. 0.25) and having more
strength at larger $y$, so that relatively more of the $\pi N$
contribution may be canceled at larger $x$ than at smaller $x$.
(Note that in covariant approaches with a $t$-dependent dipole
form factor the $\pi\Delta$ distribution is softer than the $\pi N$,
so that there the $\Delta$ plays a negligible role at large $x$.)
The cancellation of the $\bar d$ excess with the inclusion of
$\pi \Delta$ states can be seen explicitly in Fig.~7, where the
(positive) $\pi N$ and (negative) $\pi\Delta$ contributions are
shown (dashed lines) for $\Lambda_{\pi N}=1.5$~GeV and
$\Lambda_{\pi\Delta}=1.3$~GeV, together with the sum (solid).
In Fig.~7(a) the $\pi\Delta$ brings the difference $\bar d - \bar u$
closer to the large-$x$ data points, while allowing for a reasonable
fit at smaller $x$.
However, the cancellation is still not sufficient to produce a
downturn in the ratio at large $x$, as the E866 data appear to
prefer, Fig.~7(b).

More cancellation can be achieved by either increasing the
$\pi\Delta$ contribution, or decreasing the $\pi N$ contribution.
Either is acceptable in the model, as long as the resulting
distributions do not contradict other observables, such as
the total $\bar d + \bar u$ distribution, which should serve
as an absolute upper limit on the strength of the form factor
\cite{AWT83}.
In Fig.~8(a) we show the contributions to the sum $x(\bar d + \bar u)$
from the $\pi N$ and $\pi \Delta$ components with
$\Lambda_{\pi N} = 1.5$~GeV and $\Lambda_{\pi\Delta} = 1.3$~GeV,
compared with the CTEQ4 \cite{CTEQ4} and MRS98 \cite{MRST}
parameterizations.
While at small $x$ the calculated distributions lie safely below
the parameterization (the difference is made up by the perturbatively
generated $g \rightarrow q\bar q$ antiquark distributions), at large
$x$ the pion cloud already saturates the total sea with these cut-offs
--- although one should add a cautionary note that the antiquark
distribution at large $x$ is not determined very precisely.
For softer combinations of form factors, namely
$\Lambda_{\pi N} = 1$~GeV, $\Lambda_{\pi\Delta} = 1.3$~GeV and
$\Lambda_{\pi N} = \Lambda_{\pi\Delta} = 1$~GeV,
the total non-perturbative antiquark sea in Fig.~8(b) is below
the empirical parameterizations in both cases.

Therefore the only way to obtain a smaller $\bar d$ excess at
large $x$ and still be consistent with the total antiquark
distribution is to reduce the $\pi N$ component, having a
cut-off smaller than for the $\pi N \Delta$ vertex.
It was argued in Ref.\cite{E866THEORY} that the $\pi N\Delta$
form factor should be softer than the $\pi N N$, based on the
observation that the $M1$ transition form factor was softer for
$\gamma N \Delta$ than for $\gamma N N$.
However, there is no clear connection between these form factors,
and hence no compelling reason why the $\pi N \Delta$ form factor
cannot be harder than that for $\pi N N$.
Indeed, a comparison of the axial form factors for the nucleon
and for the $N$--$\Delta$ transition strongly favor an $N$--$\Delta$
axial form factor that is significantly harder than that of the nucleon.
In fact, the former is best fit by a 1.3~GeV dipole, while the latter
by a 1.02~GeV dipole parameterization \cite{AXIAL}.
Within the framework of PCAC these form factors are directly related to
the corresponding form factors for pion emission or absorption \cite{SOFT}.

In Fig.~9 we show the difference and ratio of the $\bar d$
and $\bar u$ distributions calculated with the softer $\pi N N$
form factor, $\Lambda_{\pi N} = 1$~GeV, and
$\Lambda_{\pi\Delta} = 1.3$~GeV.
The excess at large $x$ now is largely canceled by the $\pi\Delta$.
However, the smaller $\pi N$ contribution means that the asymmetry
is underestimated in the intermediate $x$ range, $x \alt 0.2$.

Based on these results it would appear difficult to obtain a
quantitative description, within the pion cloud model, of both
the ratio {\em and} the difference of the $\bar d$ and $\bar u$
distributions, together with the total sea.
One needs to consider, therefore, the possibility that other mechanisms
may at least be partly responsible for the discrepancy.
Additional meson--baryon components, such as the $\rho N$, could be
included in extended versions of the meson cloud model \cite{MTV,HSS}.
The $\rho N$, however, has a harder $y$ distribution than the $\pi N$,
which would lead to an enhancement of asymmetry at large $x$,
in contradiction with the data, even though the $\rho$ structure
function may be softer than the $\pi$.
On the other hand, the magnitude of the $\rho N$ contribution is known
from previous analyses \cite{MTV,HSS} to be significant only for very
hard $\rho N N$ form factors. For a $\rho NN$ vertex of similar shape
to that used for the $\pi NN$ vertex,
$\Lambda_{\rho N} \simeq \Lambda_{\pi N}$, the contribution
from the $\rho$ is unimportant.
Furthermore, the $\rho \Delta$ contribution is far too small to be
featured in any subsequent cancellation \cite{MTV,HSS,REVIEW},
if the $\rho \Delta$ form factor is comparable to that for $\pi \Delta$.

Another interesting possibility is that the pion sea itself could
be asymmetric.
In Eq.(\ref{conv}) only the valence structure of the pion was
included, though in principle there could even be asymmetric
contributions to $\bar d - \bar u$ in the proton from asymmetric
$\bar d^{\pi^+}$ and $\bar u^{\pi^+}$ distributions in a pion.
One obvious source of a pion sea asymmetry involves the same
physics that is responsible for the charge radius of the $\pi^+$,
namely the dissociation into virtual $\pi$ and $\rho$ mesons,
$\pi^+ \rightarrow \pi^+ \rho^0$ or $\pi^0 \rho^+$.
The effects of a meson cloud of a pion ($\pi \rho$ as well as
$\overline K K^*$ and $K \overline K^*$) on deep-inelastic structure
functions were previously investigated in Ref.\cite{PISEA}.
(Of course the $\Delta$-isobar again cancels some of this asymmetry
through the $\pi^-$, but as with the valence pion contributions,
the sign of the effect remains.)

Unfortunately, nothing is known empirically about the pion sea,
so that the shape and normalization of such an asymmetry can at
present only be speculative.
Rather than construct a detailed model of the pion sea involving
additional free parameters (meson--meson vertex functions),
at this stage it is more practical to ask how sensitive could the
overall $\bar d/\bar u$ asymmetry be to a possible asymmetric sea,
one that is consistent with all the known phenomenological constraints.
To address this question we parameterize the non-singlet part of
the sea distribution in the pion by a simple form,
\begin{eqnarray}
\bar d_{\rm sea}^{\pi^+} - \bar u_{\rm sea}^{\pi^+}
&=& {\cal N} x^\alpha (1-x)^\beta .
\end{eqnarray}
%
%
{}From low-energy quark models and Regge theory the exponent $\alpha$
is expected to be around 0 and --1/2, respectively, while $\beta$
should be between 5--7 from perturbative QCD arguments and from our
knowledge of the nucleon sea quark distribution.
The normalization of this component is unknown and given by the
parameter ${\cal N}$.
Allowing for up to a factor 2 uncertainty in the pion sea
(which is related to the uncertainty in the knowledge of
the gluon distribution in the pion), we set ${\cal N} \sim 4$,
and to be definite take $\alpha=0$ and $\beta=5$.

The resulting $\bar d/\bar u$ ratio is shown in Fig.~10 for
a ratio of pion sea distributions
$\bar d_{\rm sea}^{\pi^+} : \bar u_{\rm sea}^{\pi^+}$ = 2:1
(lower solid) and 4:1 (upper solid), respectively,
as well as for a symmetric sea (dashed) for
$\Lambda_{\pi N} = \Lambda_{\pi \Delta} = 1$ GeV.
Clearly one gets appreciable enhancement in the low- and
intermediate-$x$ range, bringing the curves to better agreement
with the data, even though the ratio is somewhat overestimated
at low $x$ for the more asymmetric sea scenario.
How reasonable this choice of parameters is can only be ascertained
by acquiring data on the pion structure function at values of $x$
smaller than currently available.
The phenomenological consequences of an asymmetric pion sea for
Drell-Yan $\pi N$ and other processes will be discussed in more
detail elsewhere \cite{PIASYM} --- see also Ref.\cite{PISEA2}
for a discussion of measurements which would be sensitive to
such an asymmetry.

Going beyond explanations involving meson clouds, one can also
investigate the possibility that the bare nucleon itself could be
asymmetric with respect to $\bar u$ and $\bar d$.
As suggested long ago by Field and Feynman \cite{FF}, the Pauli
exclusion principle can contribute to the asymmetry on the basis
of the $u$ and $d$ valence quarks being unequally represented in
the proton, thereby affecting the likelihood with which $q\bar q$
pairs can be created in different flavor channels.
In fact, earlier analyses of the NMC data \cite{SST,MTS} suggested
that the best agreement between theory and experiment could be
obtained with the combined effects of pions and antisymmetrization,
and in the next section we explore this possibility further.

\section{Antisymmetrization}

Although not directly attributable to the exclusion principle,
the perturbative effects of higher-order quark exchange diagrams
on the $\bar d - \bar u$ difference were calculated long ago by
Ross and Sachrajda \cite{RS}.
They found that while they had the correct (positive) sign,
their magnitude was insignificantly small, since they only arose
at order $\alpha_s^2$.
The flavor asymmetry of the sea associated with the Pauli principle
can therefore only be addressed within non-perturbative approaches
to parton distributions, as concluded in \cite{RS}.
Attempts to calculate the valence distribution of the proton in various
quark models began in the mid-70s \cite{EARLY}, when it was realized
that the relationship to the QCD improved parton model is quite natural
at a low scale (below 1~GeV$^2$), where most of the momentum of the
nucleon resides on its valence quarks \cite{PPNP}.
This observation has been successfully exploited by the Dortmund
group \cite{GRV}, for example, in constructing phenomenological,
valence dominated, parameterizations in just this region.

Bag model calculations of nucleon structure functions have provided some
interesting insights into the non-perturbative parton distributions
\cite{BAG,BAG1}. 
For any model in which valence quarks are confined by a strong scalar
field, the vacuum inside and outside the hadron will be different.
{}From the point of view of an external probe, such as the virtual photon
in deep inelastic scattering, the change in the vacuum structure inside
the hadron will appear as an intrinsic, non-perturbative sea of
$q \bar{q}$ pairs~\cite{BAG}.
Because of the Pauli exclusion principle, the presence of two valence
$u$ quarks, as opposed to a single valence $d$ quark, in the proton
implies an asymmetry in this non-perturbative sea, so that there is a
small excess of $d \bar{d}$ pairs over $u \bar{u}$ pairs.

{}For details of the quantitative calculation of this effect,
which is model dependent, we refer to the original papers \cite{BAG}.
It is enough for us that the shape of $\bar{d} - \bar{u}$ was found
to be similar to that of the usual sea quark distributions and the
normalization, $\int dx (\bar{d} - \bar{u})$, less than 0.25.
With this in mind, we parameterize the Pauli contribution by:
\begin{eqnarray}
\left( \bar d - \bar u \right)^{\rm Pauli}
&=& \Delta^{\rm Pauli} (n+1) (1-x)^n .
\end{eqnarray}
Because the E866 data implies a softer asymmetry than typical global
fits of total sea quark distributions would give, as Figs.~2 and 3
illustrate, phenomenologically the power $n$ should be $\agt 10$
rather than the 5--7 that has been common for the total $\bar q$ fits
\cite{CTEQ4,MRST}.
(Compare also with the original Feynman-Field parameterization \cite{FF}
which had $n = 10$ and 7 for $\bar u$ and $\bar d$, respectively.)

The Pauli effect will produce an excess of $\bar d$ over $\bar u$
over the whole range of $x$, so that it cannot lead to any cancellation
of the large-$x$ asymmetry.
To be consistent with the trend of the large-$x$ data, especially
for the $\bar d/\bar u$ ratio, one needs therefore to keep the
$\pi N N$ contribution softer than that from $\pi N \Delta$.
Taking the $\pi N$ and $\pi\Delta$ contributions calculated with
$\Lambda_{\pi N} = 1$~GeV and $\Lambda_{\pi\Delta} = 1.3$~GeV as
in Fig.~9 above, we show in Fig.~11 the combined effects of
pions and the Pauli effect.
For the latter the exponent $n = 14$, and the normalization is
$\Delta^{\rm Pauli} \approx 7\%$, which is at the lower end of the
expected scale but consistent with the bag model calculations \cite{BAG}.
Together with the integrated asymmetry from pions,
$\Delta^{\pi} \approx 0.05$, the combined value
$\Delta = \Delta^{\pi} + \Delta^{\rm Pauli} \approx 0.12$ is in quite
reasonable agreement with the experimental result, 0.100 from E866
and 0.148 from NMC.
While the quality of the fit in Fig.~11 is quite good, it would be
further improved (see Fig.~12) if one were to use the softer pion
structure function, $\bar q^\pi \sim (1-x)^2$, as suggested by
perturbative QCD \cite{FJ}.

Before leaving this discussion of antisymmetrization we should also
mention the calculations of Donoghue and Golowich \cite{DG}, and more
recently Steffens and Thomas \cite{FERN}, of the one-gluon-exchange
corrections to the 3-quark proton wave function.
Considering all possible permutations of the 5-quark wave function
allowed by Fermi statistics, the antisymmetrization of the $q\bar q$
pair split off from the emitted gluon with the quarks in the nucleon
ground state, it turns out that there are in fact more diagrams for
$u$ quarks than $d$ quarks.
This peculiarity results in there actually being an excess of $\bar u$
quarks over $\bar d$, albeit a very small one.
On the other hand, one should note that this calculation considered
just the perturbative contribution, while the Signal-Thomas effect
\cite{BAG} is a totally non-perturbative phenomenon, including all
possible non-perturbative interactions between the produced quark
(or antiquark) and the confining mean field of the proton.
Steffens and Thomas also investigated the effects of antisymmetrization
between $q\bar q$ pairs arising from one-pion loops with the three quarks
in the nucleon ground state \cite{FERN}, although here again the effects
were found to be quite small compared with the antisymmetrization for the
bare nucleon state, and from the pion cloud contribution discussed in
Section III.

\section{Conclusion}

We have, for the first time, at our disposal important new data which
map out the $x$-dependence of the asymmetry of the light antiquark sea.
Most importantly, the E866 Drell-Yan results confirm the earlier
observations that the $\bar d$ and $\bar u$ content of the proton
is not symmetric.
One of the more interesting new features of the data is the relatively
fast downturn in the $\bar d/\bar u$ ratio beyond $x \sim 0.15$,
which drops rapidly back to unity by $x \sim 0.3$.
Taken at face value, this would appear to provide a challenge to models
in which the asymmetry is assumed to arise solely from the pion cloud
of the nucleon, and in turn leads us to consider a richer and more
complex structure of the non-perturbative sea in which several mechanisms
may give competing contributions.

The evidence from the large-$x$ data indicates that a $\pi\Delta$
component in the nucleon wave function may be necessary, one which
is harder in momentum space than the $\pi N$ component.
Such a distribution arises naturally in the infinite momentum frame
formulation of the pion cloud, unlike in earlier covariant approaches
using $t$-dependent form factors where it was softer than the $\pi N$
component and hence played no role at large $x$.
Consistency with data for the sum of $\bar d$ and $\bar u$ at $x \agt 0.2$
requires that both the $\pi N N$ and $\pi N \Delta$ form factors be
relatively soft, making it difficult to avoid underestimating the E866
asymmetry at intermediate $x$, and leaving room for other effects,
such as the Pauli exclusion principle, to make up the difference.
Along the lines of previous estimates of the Pauli effect,
we find the contribution to the $\bar d-\bar u$ difference from
antisymmetrization to be significant in magnitude, and particularly
important at small $x$.
Our final results suggest that the best description of the E866 data
is indeed that in which pions and antisymmetrization play roughly
equal roles --- consistent with the findings of the earlier analysis
\cite{MTS} of the NMC data for $F_2^p - F_2^n$.

In conclusion, we note that it would be helpful to have more data at
large $x$, where the error bars are largest, to verify the downward
trend of $\bar d - \bar u$, and to further explore the possible
discrepancy between the Fermilab and CERN data.

\acknowledgements

W.M. and J.S. would like to thank the Special Research Centre for
the Subatomic Structure of Matter at the University of Adelaide for
hospitality and support.
We would also like to thank S.J. Brodsky, G.T. Garvey, E.M. Henley,
N.N. Nikolaev, J. Pumplin and F.M. Steffens for helpful discussions.
This work was supported by the Australian Research Council.


\begin{figure}		
\epsfig{figure=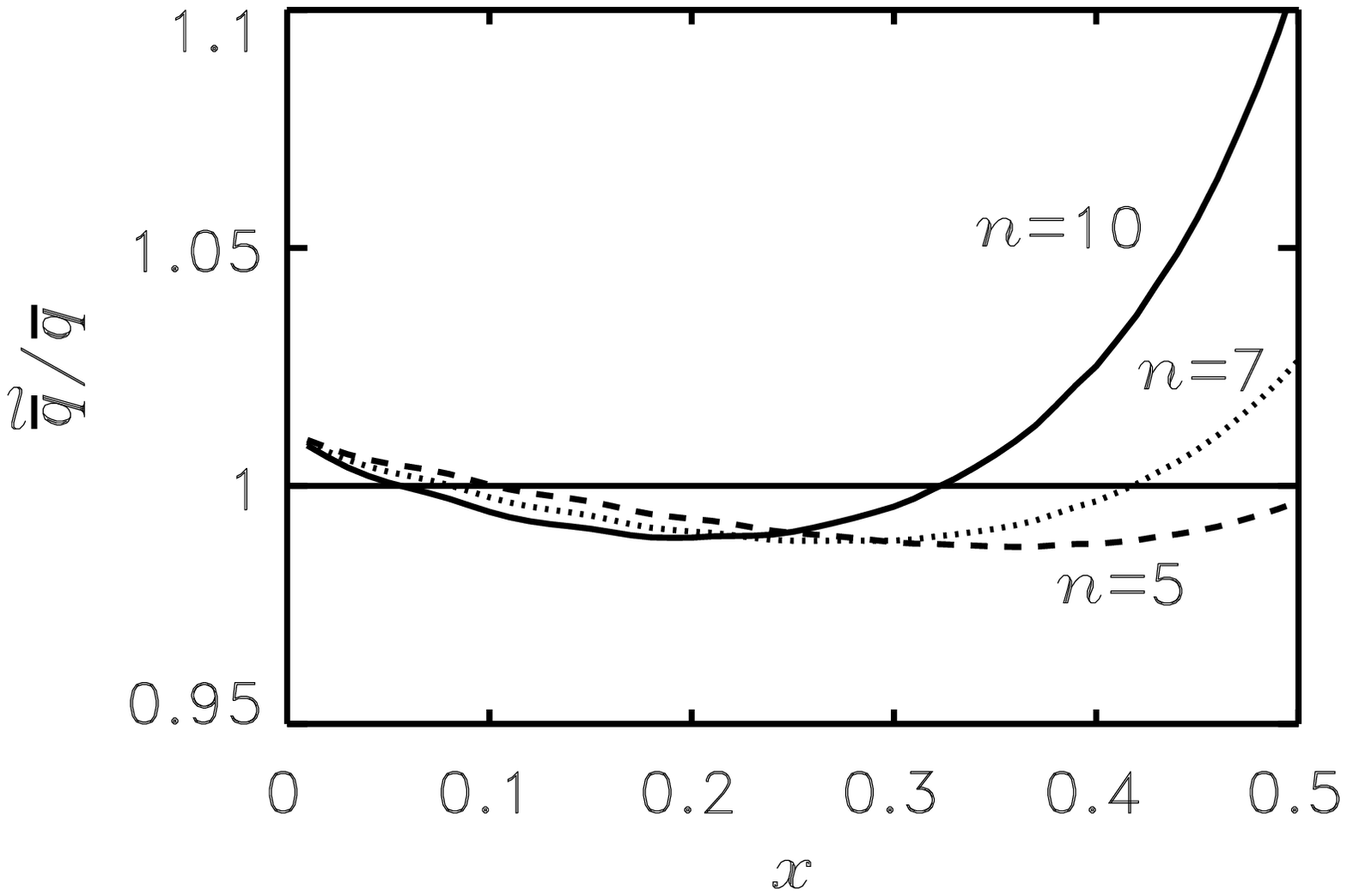,height=8.5cm}
\caption{Ratio of the antiquark distribution in a nucleon bound in the
	deuteron to that in the free nucleon, for $\bar q \sim (1-x)^n$,
	with $n=5, 7$ and 10.}
\end{figure}

\begin{figure}		
\epsfig{figure=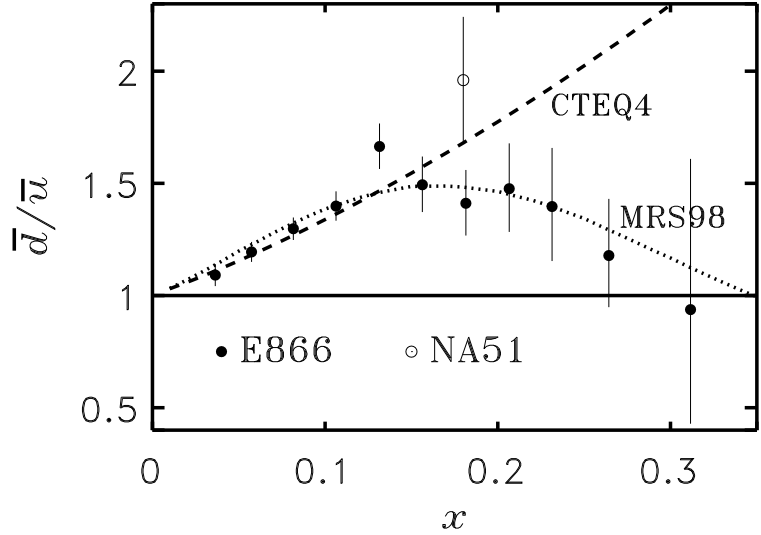,height=8.5cm}
\caption{The $x$ dependence of the $\bar d / \bar u$ ratio from
	the E866 \protect\cite{E866} (filled circles) and
	NA51 \protect\cite{NA51} (open circle) experiments,
	compared with the CTEQ4	\protect\cite{CTEQ4} and MRS98
	\protect\cite{MRST} parameterizations.
	Note that the MRS98 parameterization included the E866 data
	in their fits, while CTEQ4 predates the experiment.}
\end{figure}

\begin{figure}		
\epsfig{figure=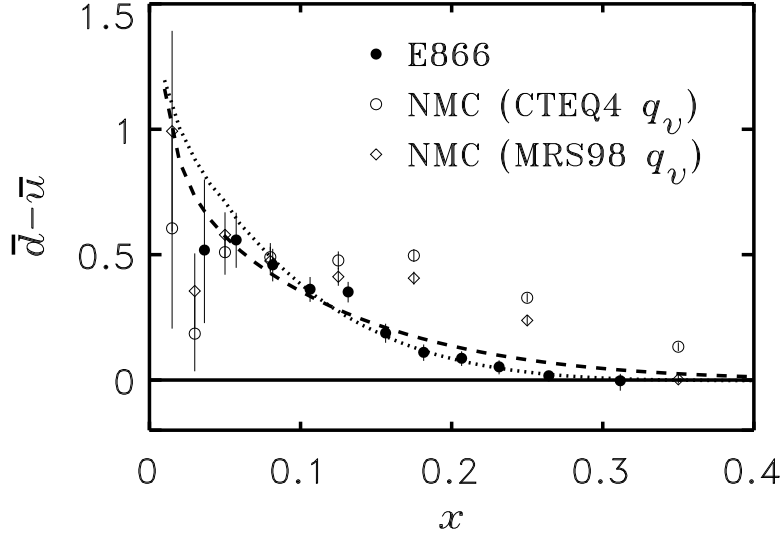,height=8.5cm}
\caption{Comparison of $\bar d - \bar u$ from the E866 experiment
	\protect\cite{E866} with the values extracted from the NMC
	measurement of the proton--neutron structure function
	difference \protect\cite{NMC}, using the CTEQ4
	\protect\cite{CTEQ4} and MRS98 \protect\cite{MRST}
	parameterizations for the valence quark distributions.
	Also shown are the parameterizations of $\bar d-\bar u$
	from CTEQ4 (dashed) and MRS98 (dotted).}
\end{figure}

\begin{figure}		
\epsfig{figure=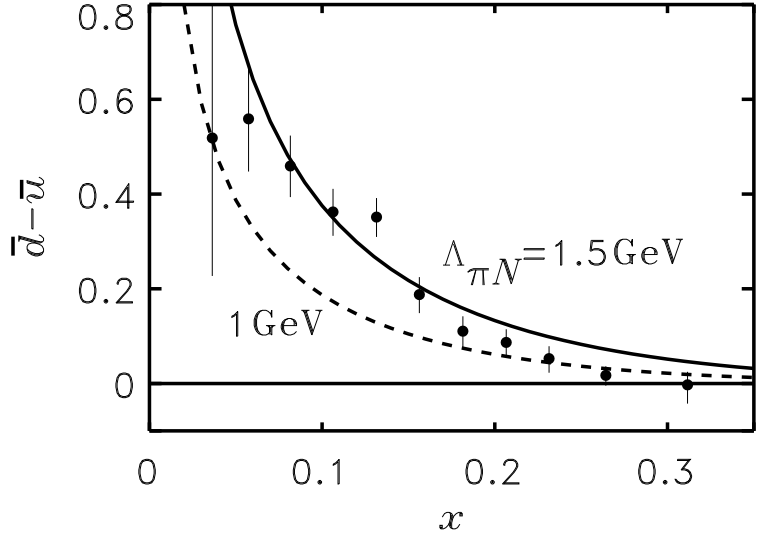,height=8.5cm}
\caption{Calculated $\bar d - \bar u$ difference arising from
	the $\pi N$ component of the proton's wave function,
	for cut-off masses $\Lambda_{\pi N}=1$~GeV (dashed)
	and 1.5~GeV (solid).}
\end{figure}

\begin{figure}		
\epsfig{figure=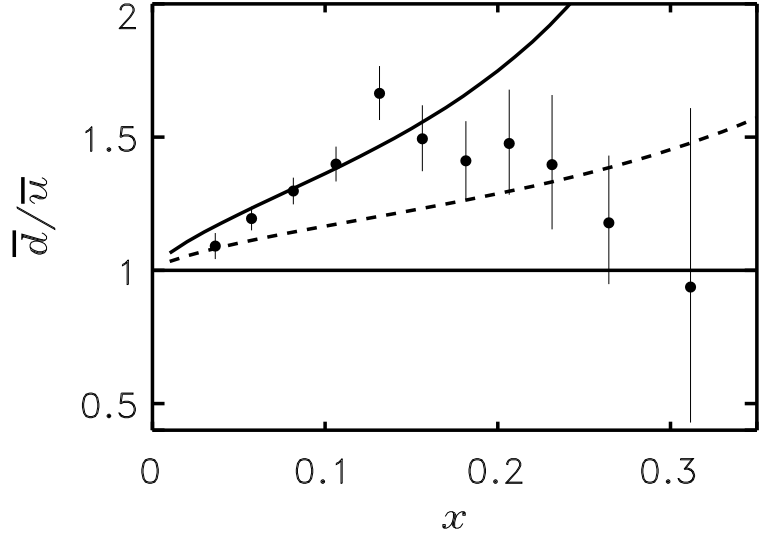,height=8.5cm}
\caption{Extracted $\bar d / \bar u$ ratio for the $\pi N$
	component, with the curves as in Fig.~4.}
\end{figure}

\begin{figure}		
\epsfig{figure=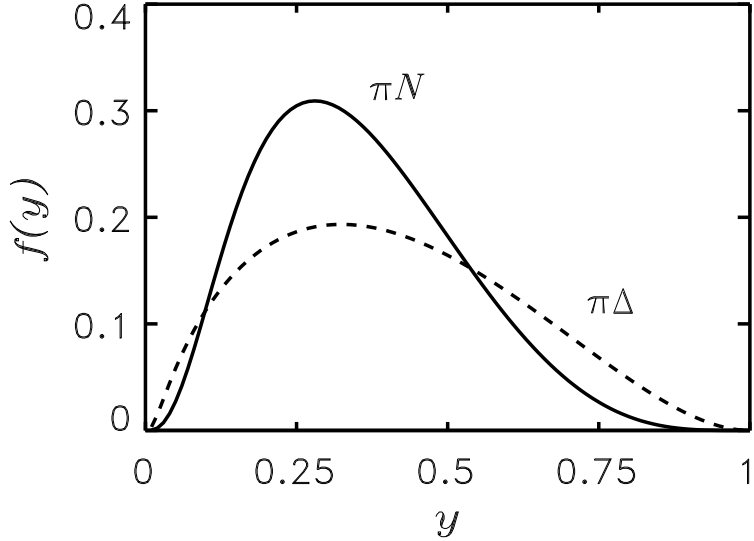,height=8.5cm}
\caption{$\pi N$ and $\pi \Delta$ momentum distribution functions,
	with dipole form factor cut-offs $\Lambda_{\pi N}=1$~GeV
	and $\Lambda_{\pi \Delta}=1.3$~GeV.}
\end{figure}

\begin{figure}		
\epsfig{figure=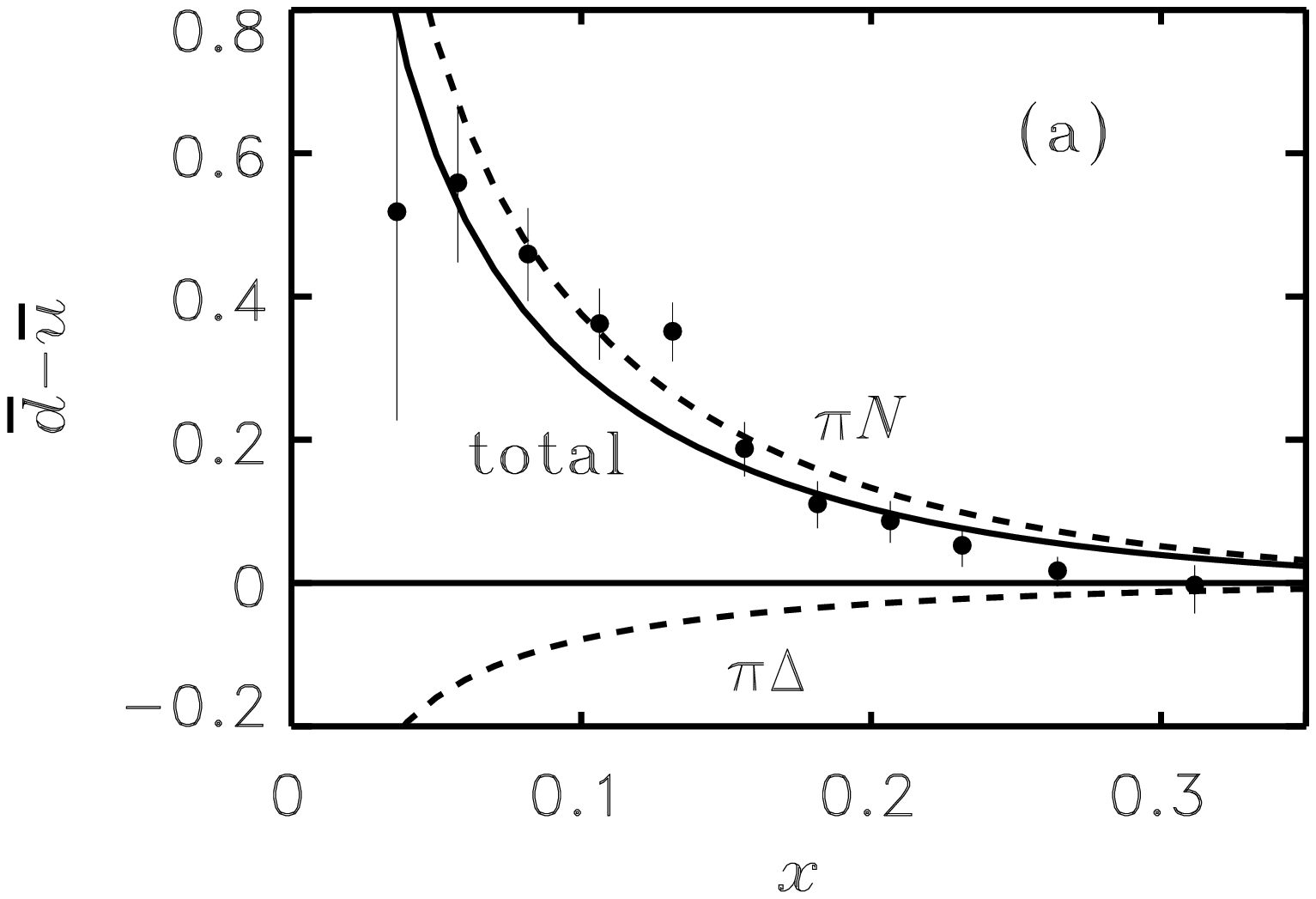,height=8.5cm}
\epsfig{figure=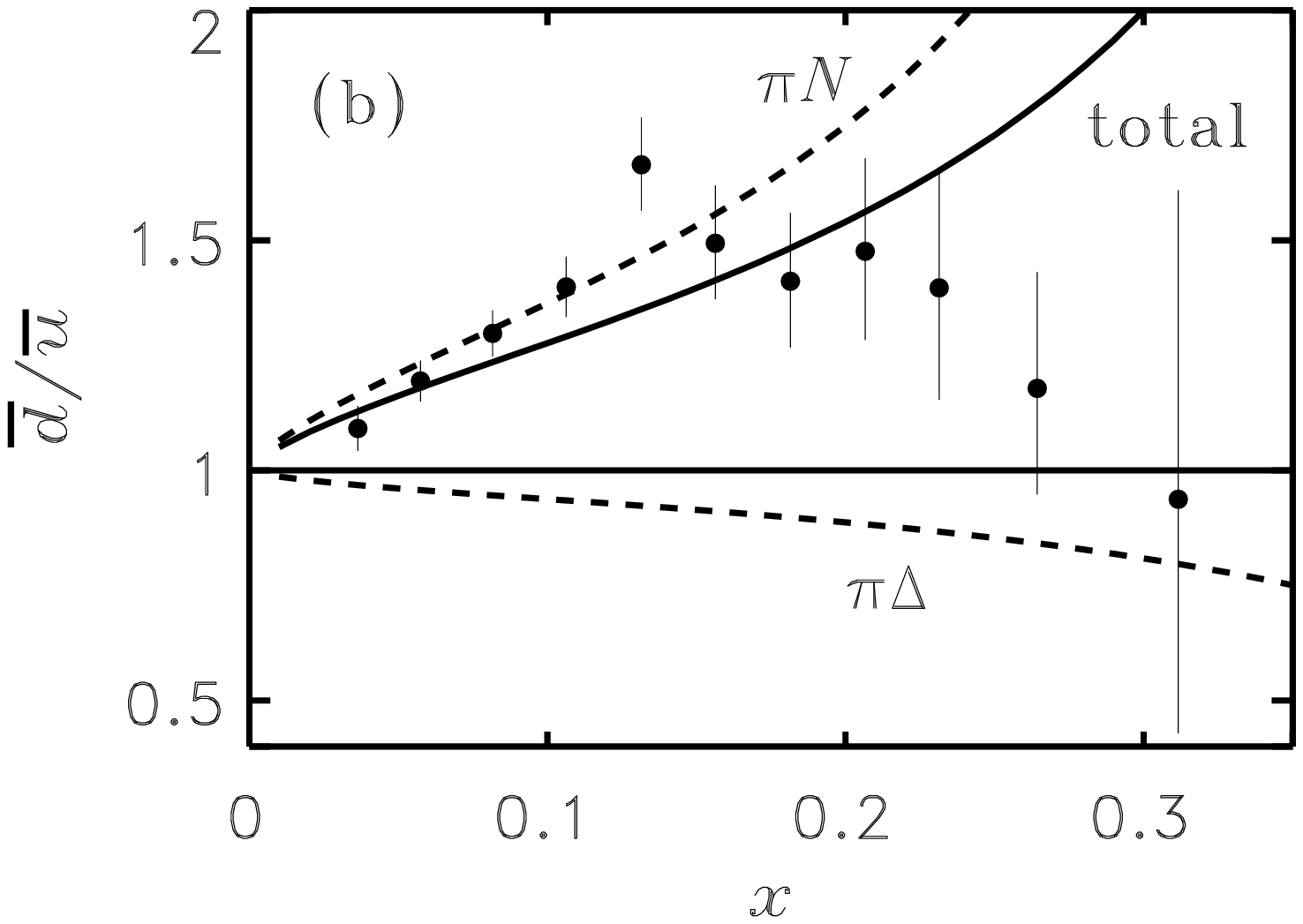,height=8.5cm}
\caption{Contributions from the $\pi N$ and $\pi\Delta$ components
	(dashed) and the combined effect (solid) to the
	(a) $\bar d - \bar u$ difference and
	(b) $\bar d/\bar u$ ratio.
	The cut-off masses are $\Lambda_{\pi N} = 1.5$~GeV
	and $\Lambda_{\pi\Delta} = 1.3$~GeV.}
\end{figure}

\begin{figure}		
\epsfig{figure=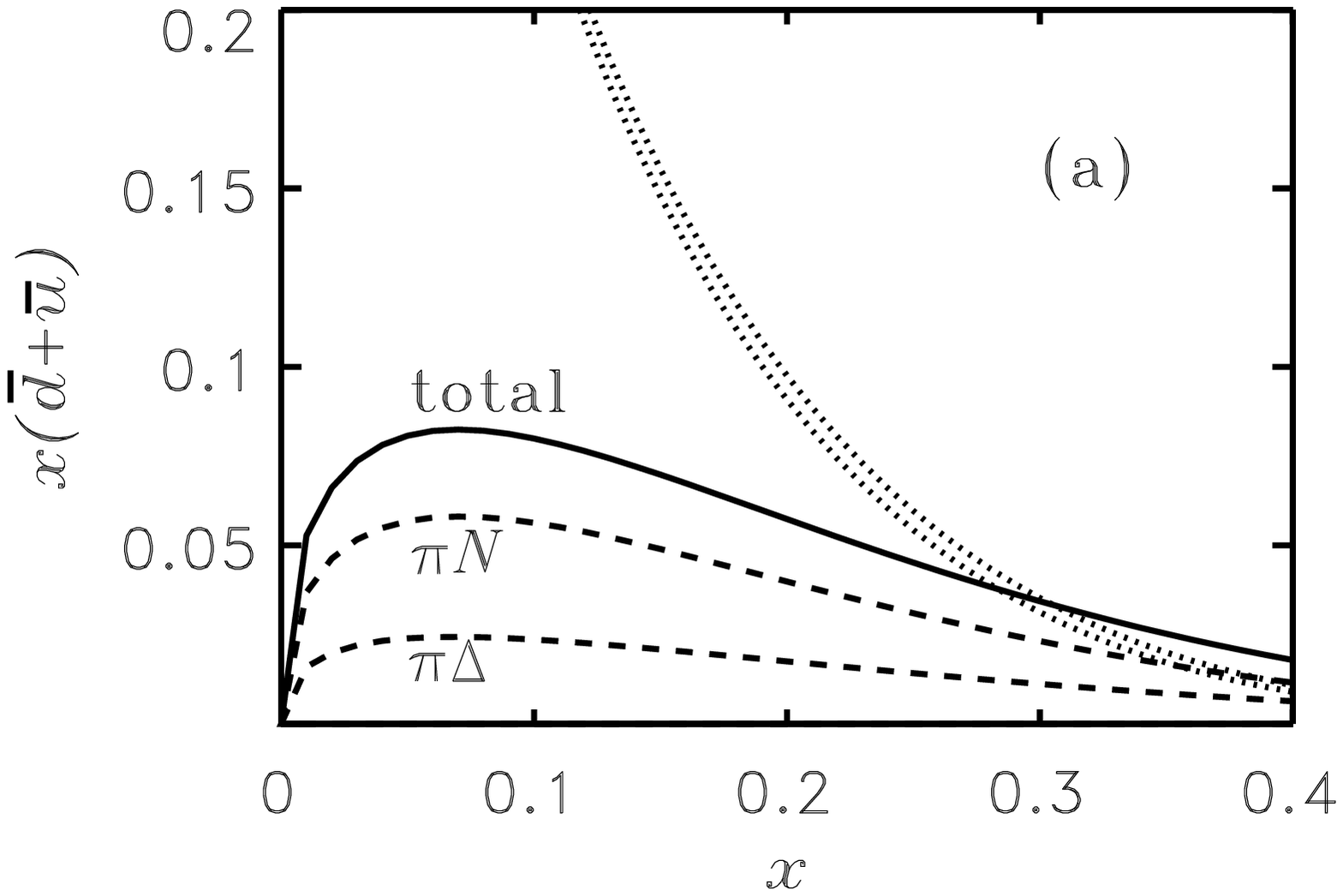,height=8.5cm}
\epsfig{figure=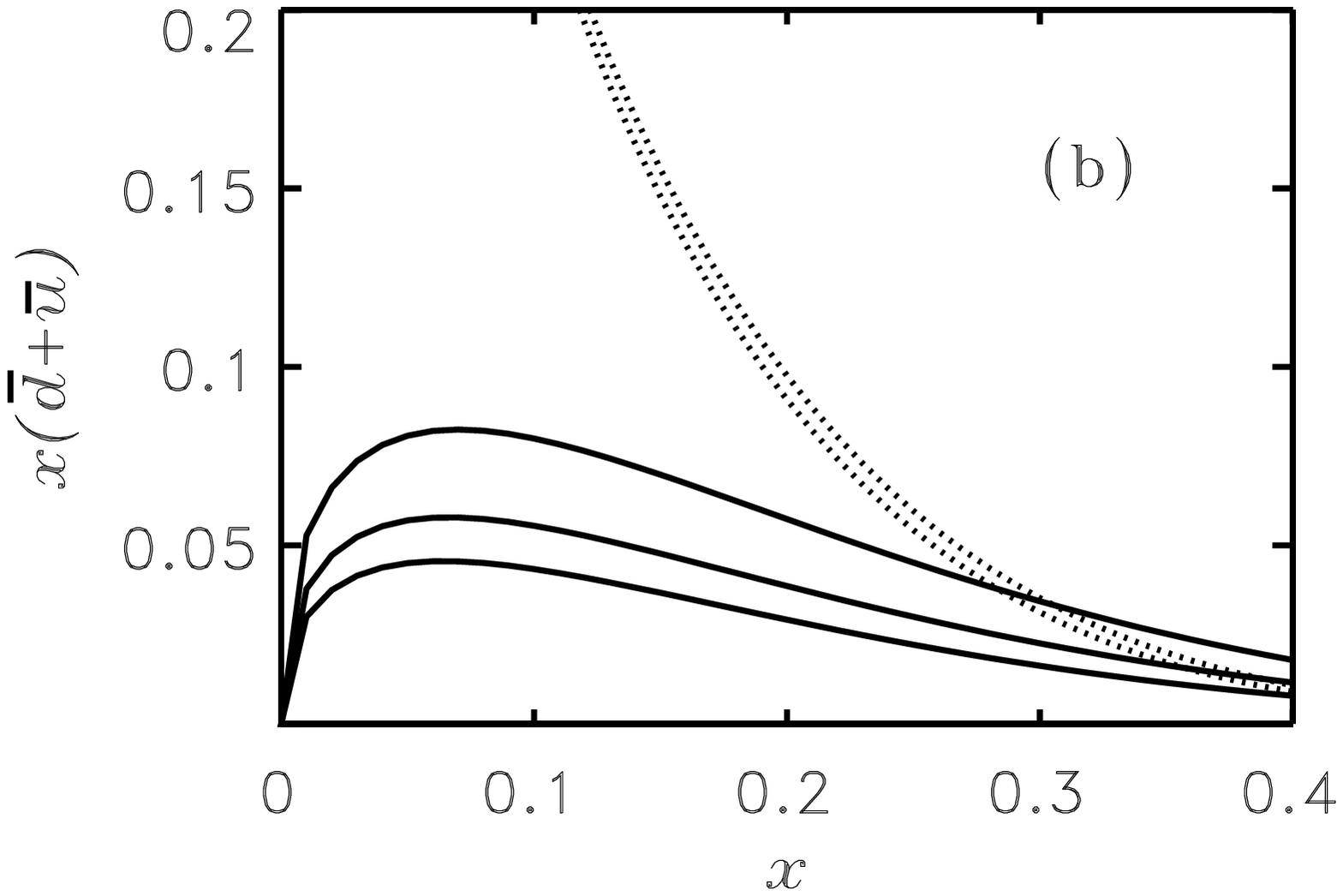,height=8.5cm}
\caption{Total $x(\bar d + \bar u)$ distribution
	(a) from the $\pi N$ and $\pi\Delta$ components (dashed), with
	$\Lambda_{\pi N} = 1.5$~GeV,
	$\Lambda_{\pi\Delta} = 1.3$~GeV,
	and the total (solid),
	(b) the total contribution for
	$\Lambda_{\pi N} = 1.5$~GeV, $\Lambda_{\pi\Delta} = 1.3$~GeV
	(largest curve),
	$\Lambda_{\pi N} = 1$~GeV, $\Lambda_{\pi\Delta} = 1.3$~GeV
	(middle), and
	$\Lambda_{\pi N} = \Lambda_{\pi\Delta} = 1$~GeV
	(smallest).
	The theoretical curves are compared with the CTEQ4
	\protect\cite{CTEQ4} and MRS98 \protect\cite{MRST}
	global parameterizations (dotted).}
\end{figure}

\begin{figure}		
\epsfig{figure=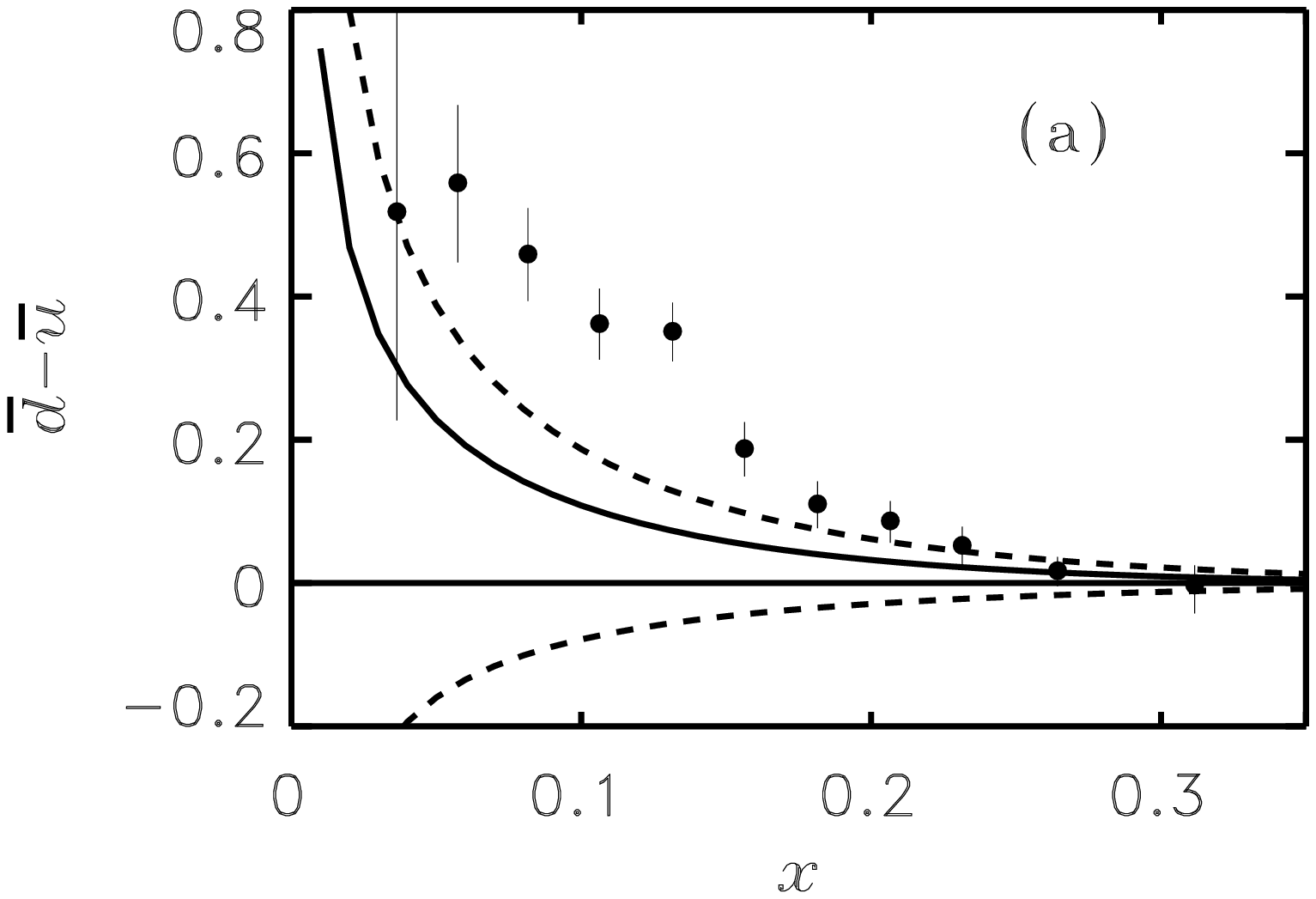,height=8.5cm}
\epsfig{figure=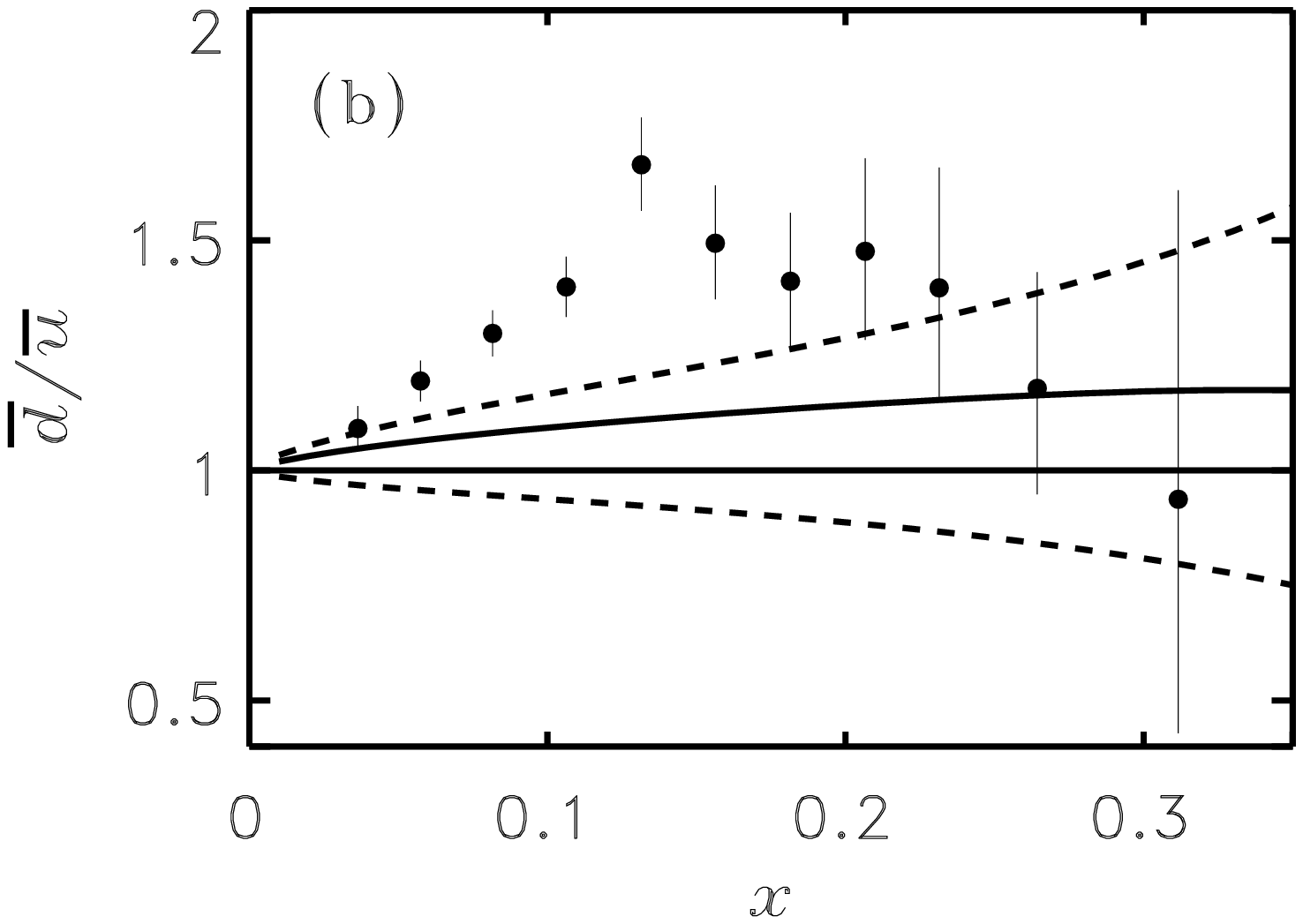,height=8.5cm}
\caption{As in Fig.7, but for $\Lambda_{\pi N} = 1$~GeV,
	$\Lambda_{\pi\Delta} = 1.3$~GeV.}
\end{figure}

\begin{figure}		
\epsfig{figure=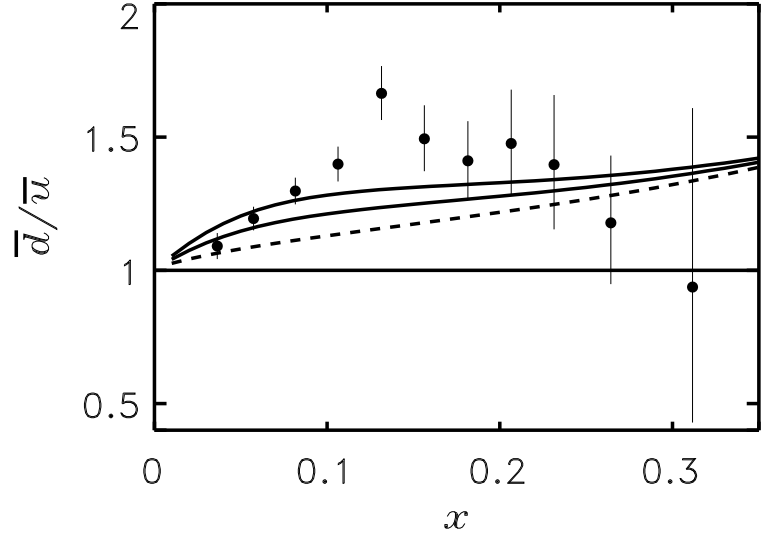,height=8.5cm}
\caption{Effect of an asymmetric pion sea on the $\bar d/\bar u$ ratio.
	The dashed curve represents the ratio for a symmetric pion sea
	with $\Lambda_{\pi N} = \Lambda_{\pi\Delta} = 1$~GeV, while
	the solid curves have asymmetric seas in the ratio
	$\bar d_{\rm sea}^{\pi^+} : \bar u_{\rm sea}^{\pi^+}
	= 2 : 1$ (lower curve) and 4 : 1 (upper curve).}
\end{figure}

\newpage

\begin{figure}		
\epsfig{figure=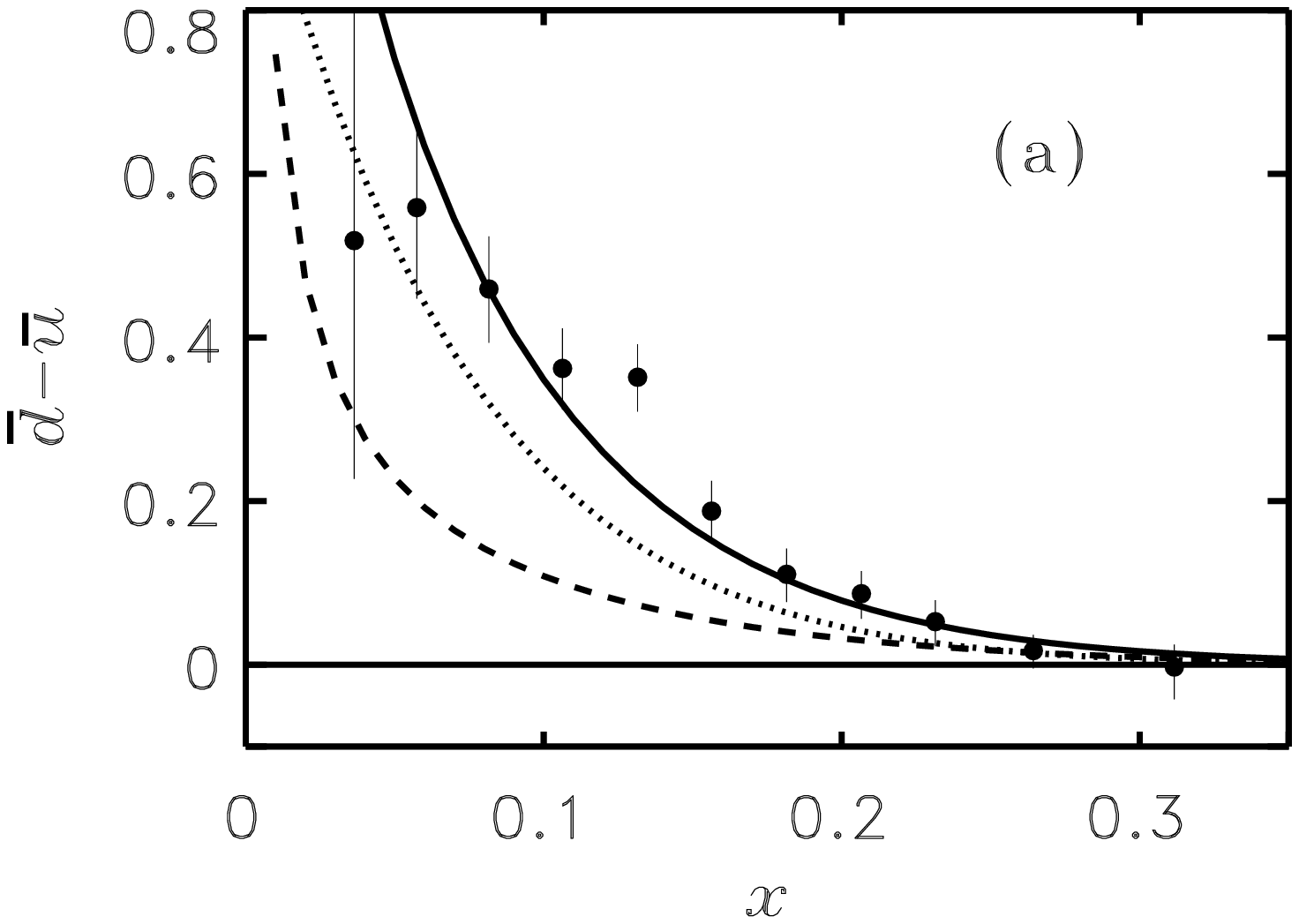,height=8.5cm}
\epsfig{figure=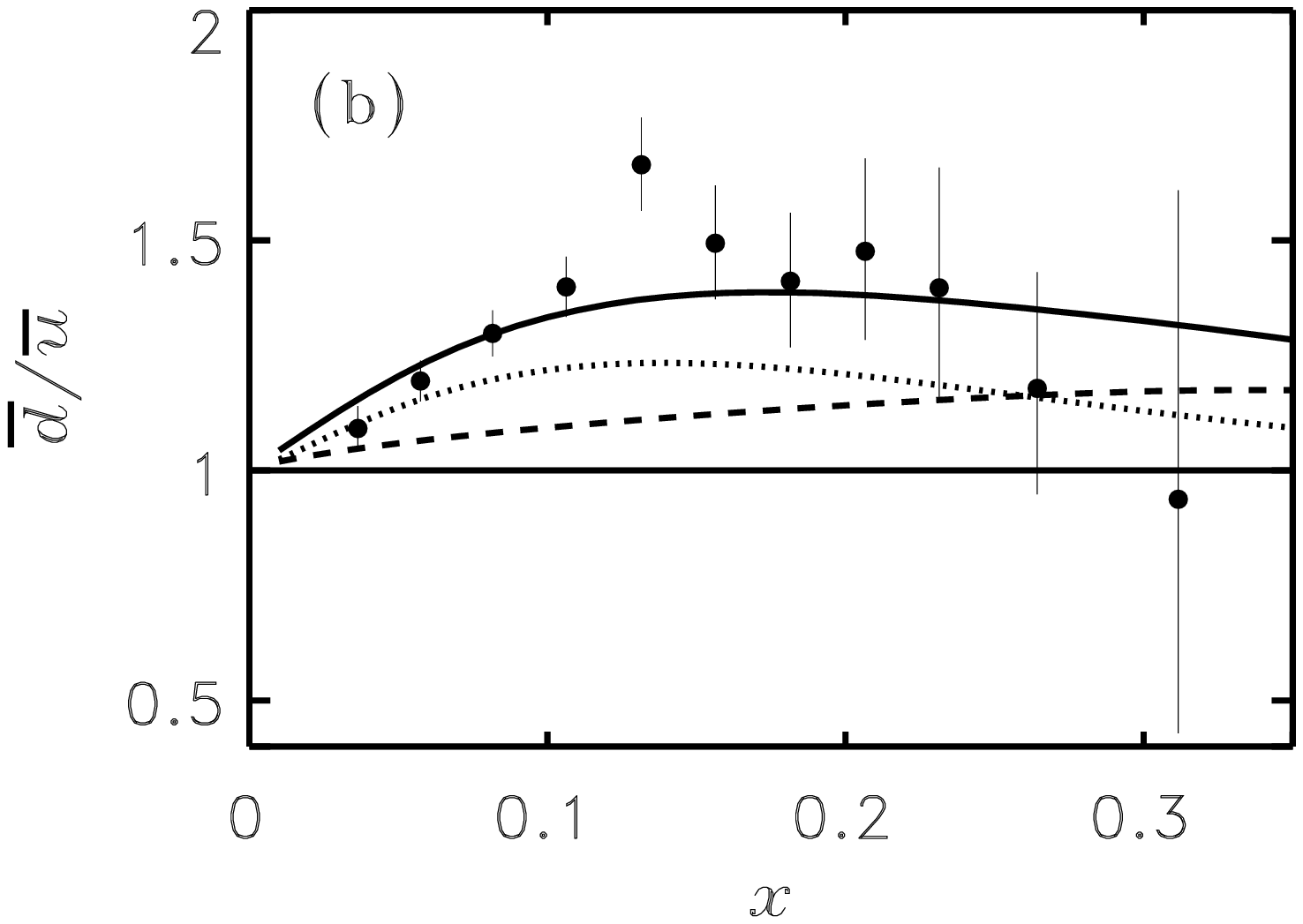,height=8.5cm}
\caption{Contributions from pions with $\Lambda_{\pi N} = 1$~GeV
	and $\Lambda_{\pi\Delta} = 1.3$~GeV (dashed) and from
	antisymmetrization (dotted) to the
	(a) $\bar d - \bar u$ difference and
	(b) $\bar d/\bar u$ ratio,
	and the combined effect (solid).}
\end{figure}

\begin{figure}		
\epsfig{figure=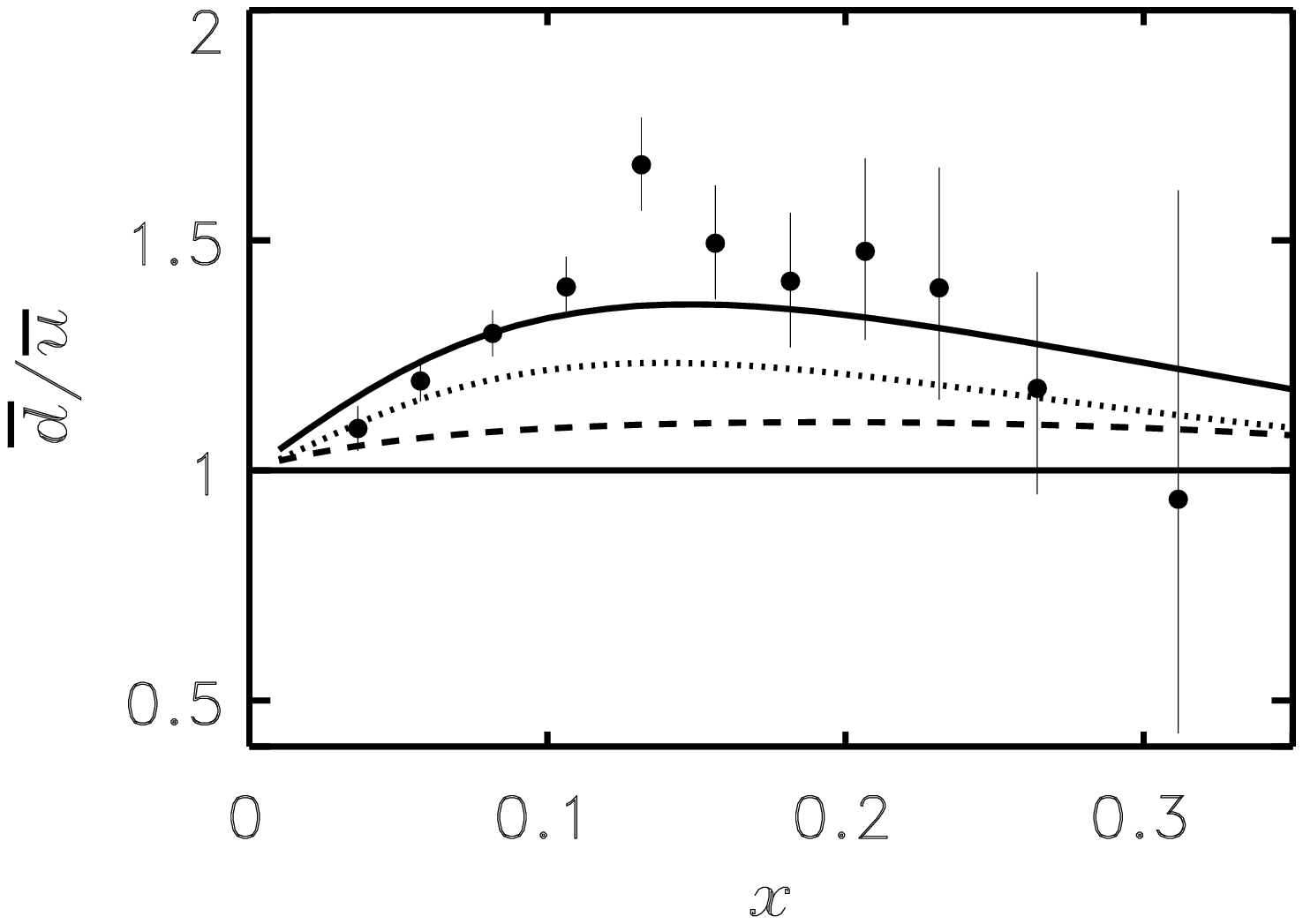,height=8.5cm}
\caption{As in Fig.11(b), but with an extra power of $(1-x)$ in the
	pion structure function, according to Ref.\protect\cite{FJ}.}
\end{figure}


\references

\bibitem{E866}
E.A. Hawker et al., E866/NuSea Collaboration,
Phys. Rev. Lett. 80 (1998) 3715.

\bibitem{NMC}
P. Amaudraz et al.,
Phys. Rev. Lett. 66 (1991) 2712;
%
M. Arneodo et al.,
Phys. Rev. D 50 (1994) R1;
%
M. Arneodo et al.,
Phys. Lett. B 364 (1995) 107.

\bibitem{NA51}
A. Baldit et al.,
Phys. Lett. B 332 (1994) 244.

\bibitem{ES}
S.D. Ellis and W.J. Stirling,
Phys. Lett. B 256 (1991) 258.

\bibitem{MTV}
A.W. Thomas and W. Melnitchouk,
in: Proceedings of the JSPS-INS Spring School, Shimoda, Japan
(eds. S. Homma et al., World Scientific, Singapore, 1993);
%
W. Melnitchouk and A.W. Thomas,
Phys. Rev. D 47 (1993) 3794.

\bibitem{AWT83}
A.W. Thomas,
Phys. Lett. B 126 (1983) 97.

\bibitem{FF}
R.D. Field and R.P. Feynman,
Phys. Rev. D 15 (1977) 2590.

\bibitem{BAG}
A.I. Signal and A.W. Thomas,
Phys. Lett. B 211 (1988) 481;
%
A.I. Signal and A.W. Thomas,
Phys. Rev. D 40 (1989) 2832;
%
A.W. Schreiber, A.I. Signal and A.W. Thomas,
Phys. Rev. D 44 (1991) 2653.

\bibitem{SST}
A.I. Signal, A.W. Schreiber and A.W. Thomas,
Mod. Phys. Lett. A 6 (1991) 271.

\bibitem{MTS}
W. Melnitchouk, A.W. Thomas and A.I. Signal,
Z. Phys. A 340 (1991) 85.

\bibitem{SMEAR}
W. Melnitchouk, A.W. Schreiber and A.W. Thomas,
Phys. Lett. B 335 (1994) 11.

\bibitem{SMEAR2}
A. Bodek and J.L. Ritchie,
Phys. Rev. D 23 (1981) 1070;
%
G.V. Dunne and A.W. Thomas,
Nucl. Phys. A455 (1986) 701;
%
L.P. Kaptari and A.Yu. Umnikov,
Phys. Lett. B 259 (1991) 155;
%
M.A. Braun and M.V. Tokarev,
Phys. Lett. B 320 (1994) 381.

\bibitem{MTD}
W. Melnitchouk and A.W. Thomas,
Phys. Rev. D 47 (1993) 3783.

\bibitem{NP}
W. Melnitchouk and A.W. Thomas,
Phys. Lett. B 377 (1996) 11.

\bibitem{CTEQ4}
H.L. Lai, J. Huston, S. Kuhlmann, F. Olness, J. Owens,
D. Soper, W.K. Tung and H. Weerts,
Phys. Rev. D 55 (1997) 1280.

\bibitem{MRST}
A.D. Martin, R.G. Roberts, W.J. Stirling and R.S. Thorne,
Eur. Phys. J. C 4 (1998) 463.

\bibitem{E866THEORY}
J.C. Peng et al., E866/NuSea Collaboration,
hep-ph/9804288.

\bibitem{RS}
D.A. Ross and C.T. Sachrajda,
Nucl. Phys. B 149 (1979) 497.

\bibitem{SULL}
J.D. Sullivan,
Phys. Rev. D 5 (1972) 1732.

\bibitem{EMCEFFECT}
J.J. Aubert et al.,
Phys. Lett. 123 B (1983) 275.

\bibitem{HM}
E.M. Henley and G.A. Miller,
Phys. Lett. B 251 (1990) 453.

\bibitem{KL}
S. Kumano,
Phys. Rev. D 43 (1991) 3067;
%
S. Kumano and J.T. Londergan,
Phys. Rev. D 44 (1991) 717.

\bibitem{HSB}
W.-Y.P. Hwang, J. Speth and G.E. Brown,
Z. Phys. A 339 (1991) 383.

\bibitem{HSS}
H. Holtmann, A. Szczurek and J. Speth,
Nucl. Phys. A569 (1996) 631.

\bibitem{EHQ}
E.J. Eichten, I. Hinchliffe and C. Quigg,
Phys. Rev. D 47 (1993) R747.

\bibitem{KFS}
W. Koepf, L.L. Frankfurt and M.I. Strikman,
Phys. Rev. D 53 (1996) 2586.

\bibitem{REVIEW}
J. Speth and A.W. Thomas,
Adv. Nucl. Phys. 24 (1998) 83.

\bibitem{MM}
W. Melnitchouk and M. Malheiro,
Phys. Rev. C 55 (1997) 431;
%
M. Malheiro and W. Melnitchouk,
Phys. Rev. C 56 (1997) 2373.

\bibitem{CHARM}
W. Melnitchouk and A.W. Thomas,
Phys. Lett. B 414 (1997) 134.

\bibitem{PNNDB}
S. Paiva, M. Nielsen, F.S. Navarra, F.O. Duraes, L.L. Barz,
Sao Paulo U. preprint IFUSP-P-1240,
hep-ph/9610310.

\bibitem{ZOL}
V.R. Zoller,
Z. Phys. C 54 (1992) 425;
{\em ibid} C 60 (1993) 141.

\bibitem{NSZ}
N.N. Nikolaev, J. Speth and B.G. Zakharov,
J\"ulich preprint KFA-IKP-TH-1997-17,
hep-ph/9708290.

\bibitem{E615}
J.S. Conway et al.,
Phys. Rev. D 39 (1989) 92.

\bibitem{NA10}
B. Betev et al.,
Z. Phys. C 28 (1985) 15.

\bibitem{NA3}
J. Badier et al.,
Z. Phys. C 18 (1983) 281.

\bibitem{GRSPI}
M. Gluck, E. Reya and M. Stratmann,
Eur. Phys. J. C 2 (1998) 159.

\bibitem{SMRS}
P.J. Sutton, A.D. Martin, R.G. Roberts and W.J. Stirling,
Phys. Rev. D 45 (1992) 2349.

\bibitem{FJ}
G.R. Farrar and D.R. Jackson,
Phys. Rev. Lett. 35 (1975) 1416.

\bibitem{BBS}
S.J. Brodsky, M. Burkardt and I. Schmidt,
Nucl. Phys. B441 (1995) 197.

\bibitem{PUMPLIN}
J. Pumplin,
Phys. Rev. D 8 (1973) 2249.

\bibitem{PI_N}
S. Th\'eberge, G.A. Miller and A.W. Thomas,
Can. J. Phys. 60 (1982) 59;
%
I.R. Afnan and B.C. Pearce,
Phys. Rev. C 35 (1987) 737;
%
C. Sch\"utz, J.W. Durso, K. Holinde and J. Speth,
Phys. Rev. C 49 (1994) 2671;
%
C. Sch\"utz, J. Haidenbauer, J. Speth and J.W. Durso,
Phys. Rev. C 57 (1998) 1464.

\bibitem{AXIAL}
G.T. Jones et al.,
Z. Phys. C 43 (1989) 527;
%
T. Kitagaki et al.,
Phys. Rev. D 42 (1990) 1331.

\bibitem{SOFT}
K. Holinde and A.W. Thomas,
Phys. Rev. C 42 (1990) 1195;
%
J. Haidenbauer, K. Holinde and A.W. Thomas,
Few-Body (ICFBP 14) (1994) 490.

\bibitem{PISEA}
W.-Y.P. Hwang and J. Speth,
Phys. Rev. D 45 (1992) 3061.

\bibitem{PIASYM}
W. Melnitchouk, J. Speth and A.W. Thomas,
in preparation.

\bibitem{PISEA2}
J.T. Londergan et al.,
Phys. Lett. B 361 (1995) 110.

\bibitem{EARLY}
A. Le Yaouanc et al.,
Phys. Rev. D 11 (1975) 680;
%
G. Parisi and R. Petronzio,
Phys. Lett. B 62 (1976) 331;
%
R.L. Jaffe,
Phys. Rev. D 11 (1975) 1953;
%
J. Bell,
Phys. Lett. B 74 (1978) 77.

\bibitem{PPNP}
A.W. Thomas,
Prog. Part. Nucl. Phys. 20 (1988) 21.

\bibitem{GRV}
M. Gluck, E. Reya, A. Vogt,
Z. Phys. C 48 (1990) 471.

\bibitem{BAG1}
A.W. Schreiber, P.J. Mulders, A.I. Signal and A.W. Thomas,
Phys. Rev. D 45 (1992) 3069;
%
F.M. Steffens, H. Holtmann and A.W. Thomas,
Phys. Lett. B 358 (1995) 139.

\bibitem{DG}
J.F. Donoghue and E. Golowich,
Phys. Rev. D 15 (1977) 3421.

\bibitem{FERN}
F.M. Steffens and A.W. Thomas,
Phys. Rev. C 55 (1997) 900.
\end{document}